\documentclass[nohyperref]{article}

\usepackage{microtype}
\usepackage{graphicx}
\usepackage{subfigure}
\usepackage{booktabs} 
\usepackage{bbm}
\usepackage{multicol}
\usepackage{multirow}
\usepackage{hyperref}

\usepackage[accepted]{icml2023}

\usepackage{amsmath}
\usepackage{amssymb}
\usepackage{mathtools}
\usepackage{amsthm}

\usepackage[capitalize,noabbrev]{cleveref}

\theoremstyle{plain}

\theoremstyle{definition}

\theoremstyle{remark}

\usepackage[textsize=tiny]{todonotes}

\icmltitlerunning{Beyond Homophily: Reconstructing Structure for Graph-agnostic Clustering}

\begin{document}

\twocolumn[
\icmltitle{Beyond Homophily: Reconstructing Structure for Graph-agnostic Clustering }




\icmlsetsymbol{corre}{*}

\begin{icmlauthorlist}
\icmlauthor{Erlin Pan}{sch}
\icmlauthor{Zhao Kang}{sch,corre}

\end{icmlauthorlist}


\icmlaffiliation{sch}{School of Computer Science and Engineering,  University of Electronic Science and Technology of China, Chengdu, China}

\icmlcorrespondingauthor{Erlin Pan}{wujisixsix6@gmail.com}
\icmlcorrespondingauthor{Zhao Kang}{ zkang@uestc.edu.cn}

\icmlkeywords{Unsupervised graph representation learning, heterophily}

\vskip 0.3in
]



\printAffiliationsAndNotice{\textsuperscript{*}Corresponding author\quad} 


	\begin{abstract}
		Graph neural networks (GNNs) based methods have achieved impressive performance on node clustering task. However, they are designed on the homophilic assumption of graph and clustering on heterophilic graph is  overlooked. Due to the lack of labels, it is impossible to first identify a graph as homophilic or heterophilic before a suitable GNN model can be found. Hence, clustering on real-world graph with various levels of homophily poses a new challenge to the graph research community. To fill this gap, we propose a novel graph clustering method, which contains three key components: graph reconstruction, a mixed filter, and dual graph clustering network. To be graph-agnostic, we empirically construct two graphs which are high homophily and heterophily from each data. The mixed filter based on the new graphs extracts both low-frequency and high-frequency information. To reduce the adverse coupling between node attribute and topological structure, we separately map them into two subspaces in dual graph clustering network. Extensive experiments on 11 benchmark graphs demonstrate our promising performance. In particular, our method dominates others on heterophilic graphs. The code is available at \href{https://github.com/Panern/DGCN}{DGCN}.
	\end{abstract}
 \section{Introduction}
	Graphs are pervasive and have been widely used in numerous real-world scenarios, such as social networks, traffic networks, and recommendation systems \cite{liu2021self}. Graph clustering that groups nodes without the need of human annotation is a fundamental yet challenging graph analysis task \cite{AGC,liu2022multilayer}. Based on GNN' powerful structure information exploitation capability, many clustering methods \cite{DAEGC, O2MAC} have achieved remarkable performance on homophilic
	graphs.\\
	However, there are many heterophilic graphs in which most of connected nodes belong to different classes \cite{Geom-GCN, xie2023contrastive}. Traditional GNNs learn  representations via message passing mechanism under the assumption of homophily \cite{fang2022structure}. Facing heterophilic graphs, previous approaches mainly suffer two limitations. On the one hand, the local neighbors in a graph are nodes that are proximally located, while nodes that are semantically similar might be far apart on
heterophilic graph \cite{H2GCN}. Thus, existing techniques fail to capture long-range information from distant nodes. On the other hand, they don't distinguish similar and dissimilar neighbors, which carry different amounts of information. Learning a discriminative graph representation needs to pass diverse messages between nodes on heterophilic graphs. 
  Consequently, the GNN-based methods performing well on homophilic graphs produce unsatisfactory performance on heterophilic graphs \cite{abu2019mixhop}.\\
	Numerous GNN methods have been proposed to deal with heterophily. Some approaches expand neighbor fields, while others refine the GNN architectures. The first class includes exploring high-order information \cite{TDGCN} and mining more neighbors in other spaces \cite{HOG-GNN}. The latter class aggregates message adaptively, like adaptive filter \cite{BernNet, ACMNN, JacobiNet} and attention or weight mechanism \cite{FAGCN,DMP}. Moreover, several recent methods, like \cite{DHGR} and \cite{SDRF}, preprocess heterophily graphs to make them fit for GNNs.\\
	Though the aforementioned methods improve the performance of GNNs on heterophilic graphs in some downstream tasks, there exists two critical problems: 1) the training of customized network, the learning of adaptive filter, and graph rewiring \cite{DHGR} rely on labelled samples, which makes them not be applicable to clustering task. 2) GNNs embed the raw data into a single subspace, where the coupling of attribute and topological structure aggravates the adverse effect of heterophily. Any incorrect or incomplete in the attributes or structures  would deteriorate the quality of learned representations.\\
 \textbf{For unsupervised learning, the first and foremost challenge we face is that there is no labels for us to judge whether a graph is homophilic or heterophilic.} Therefore, it is not practical to develop individual models to handle homophilic and heterophilic graphs separately for clustering. Moreover, it is also subjective to simply classify a graph as homophilic or
heterophilic since real-world graph data could have various levels of homophily.
	To address this open problem,  we propose a holistic approach in this work to handle real graphs, which includes three key components: graph reconstruction, a mixed filter, and dual graph clustering network. We first construct new homophilic and heterophilic graphs to explore both low-frequency and high-frequency information. In particular, the structure reconstruction process is fully unsupervised and general. The mixed filter is designed to smooth graph signals, which makes our model be applicable to both homophilic and heterophilic graphs.  Finally, the smoothed features are fed into dual graph clustering network to obtain the clustering result. We summarize our contributions as follows:
	\begin{itemize}
 	\item We propose two unsupervised graph construction strategies to extract homophilic and heterophilic information from any type of graphs.
		\item We design a mixed filter that exploits both low-frequency and high-frequency components of data. This approach can also be applied to classification task.
	
		\item We reduce the adverse coupling between attribute and topological structure by mapping them into two different subspaces.
		\item Extensive experiments on homophilic and heterophilic graphs  demonstrate the promising performance of our proposed method. 
	\end{itemize}

	\section{Related Work}
	\subsection{ Graph Clustering}
	Numerous attributed graph clustering methods  have been proposed to exploit nodes’ feature and topological structure information. These methods can be roughly classified into two categories: GNN-based methods and shallow graph embeddings based methods. GNN-based methods learn the graph representation for clustering via aggregating neighborhood information in prior graph \cite{GAE}. To improve the clustering performance, \cite{MAGCN} and \cite{AGN-H} adopt attention mechanism to adaptively integrate topology and  attribute information. Inspired by the success of contrastive learning, \cite{SAGC} learns a consensus representation from multiview graph.  Shallow methods learn graph embeddings without neural networks and perform traditional clustering methods on them. For example, \cite{AGC} obtains more discriminative representations by enlarging receptive field to explore high-order information. \cite{MvAGC} and \cite{lin2023multi} learn clustering-favorable embeddings via low-pass filter. \cite{MCGC} constructs a new graph for clustering via pulling nearest neighbors close. \\
    However, these methods only focus on homophilic graphs and are not directly applicable to heterophilic graphs. In this work, we aim to develop an omnipotent method, which is suitable for real graph with different levels of homophily.

	\subsection{Heterophilic Graph Learning}
    Heterophilic structure is prevalent in practice, from personal relationships in daily life to chemical and molecular scientific study. Developing powerful heterophilic GNN models is a hot research topic. \cite{lim1, lim2} provide general benchmarks for heterophilic graph learning. In addition, many methods have been proposed to revise GNNs for heterophilic graphs. \cite{DMP} specifies propagation weight for each attribute to make GNNs fit heterophilic graphs and \cite{GloGNN} explores the underlying homophilic information by capturing the global correlation of nodes. \cite{H2GCN} enlarges receptive field via exploring high-order structure. \cite{GPR} adaptively combines the representation of each layer and \cite{GCNII} integrates embeddings from different depths with residual operation. \\
    Although these approaches alleviate the heterophilic problem to some extent, they rely on prior knowledge like labels for  training, which are not available in unsupervised scenario. To our best knowledge, clustering on heterophilic graph has never been investigated. To fill this gap, we reconstruct homophilic and heterophilic graphs to make the proposed model handle any kinds of graphs. 
 \begin{figure*}[t]
		\centering
		\subfigure[]{
			\includegraphics[width=0.31\linewidth]{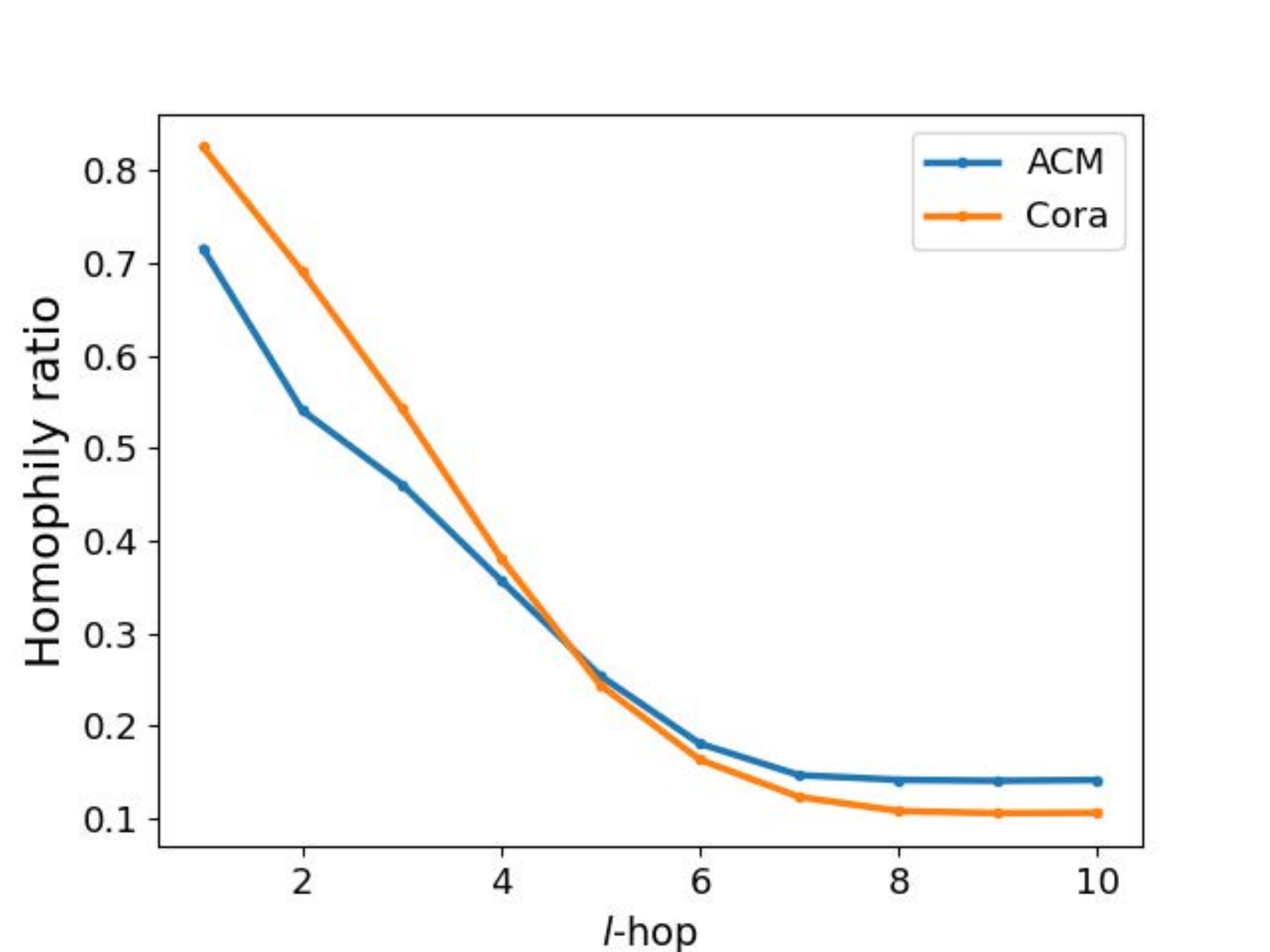}
		}
		\subfigure[]{
			\includegraphics[width=0.31\linewidth]{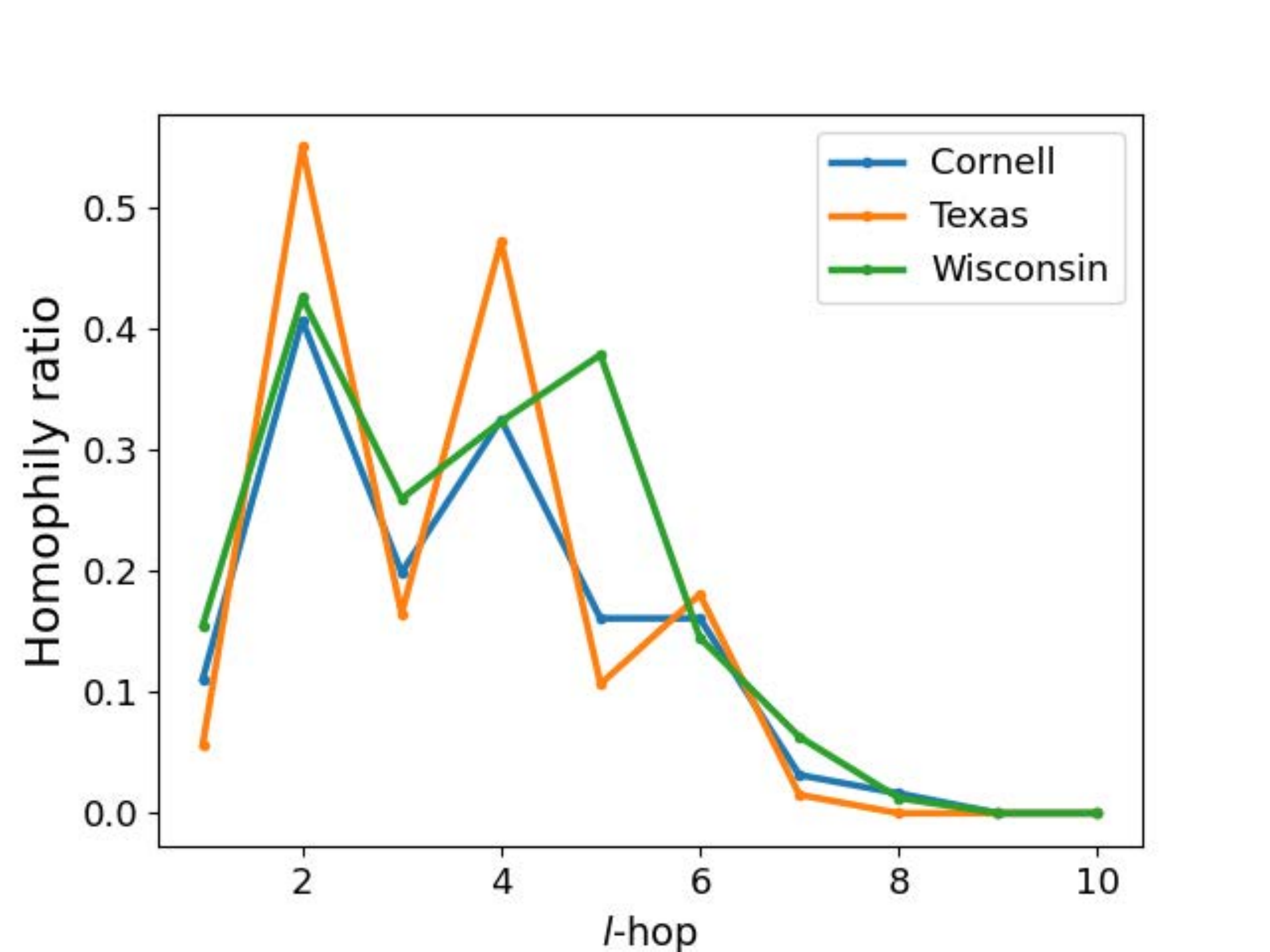}
		}
		\subfigure[]{
			\includegraphics[width=0.31\linewidth]{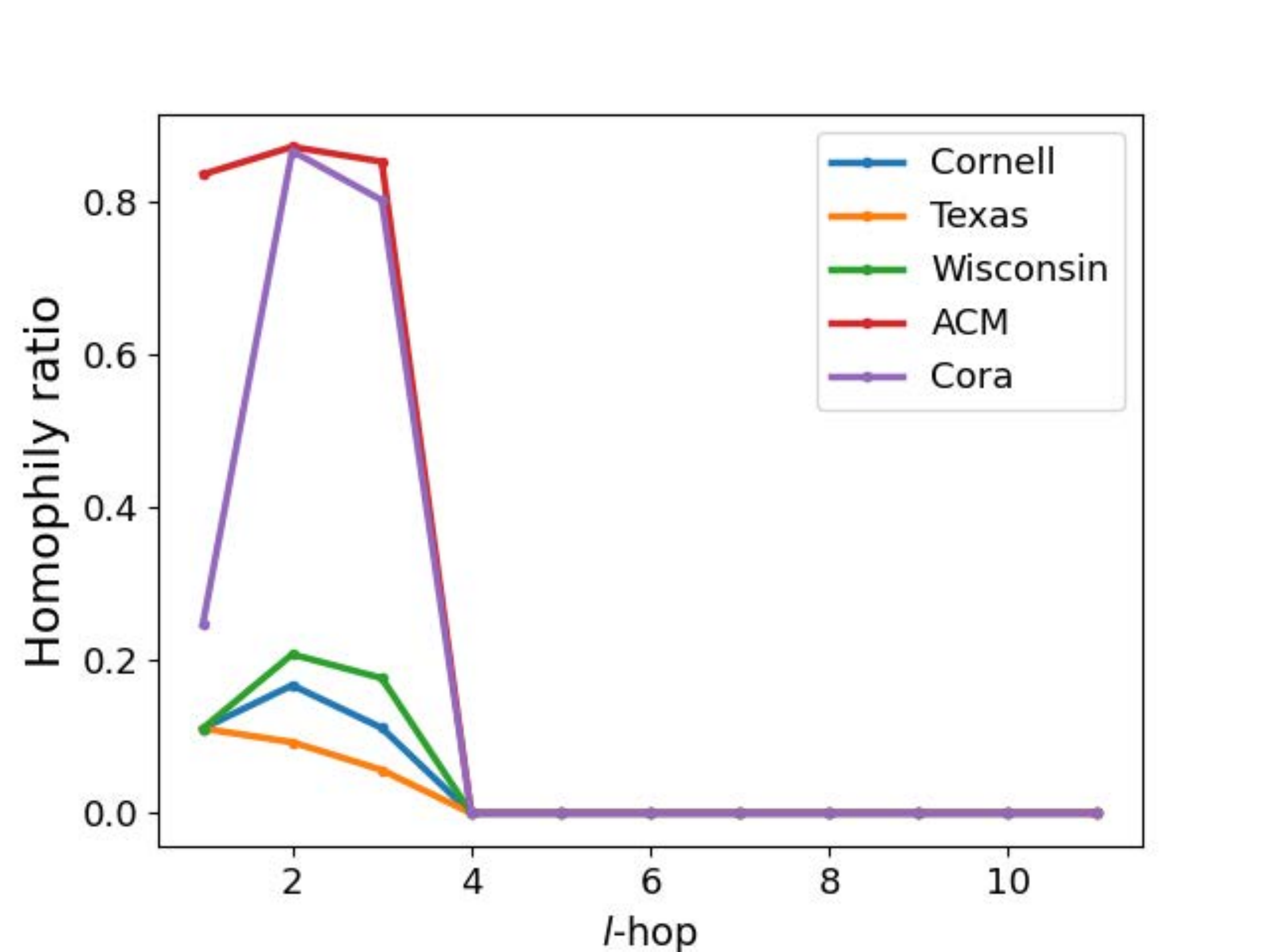}
		}
            \caption{(a) and (b) show the changes of homophily ratio in homophilic and heterophilic graphs. (c) shows the homophily ratios of 1-hop and other hops nodes sharing the same label. It's clear that homophily ratio of homophilic graphs decreases when more hops are considered. However, the change in heterophilic graphs is irregular. Particularly, the nodes in 1-hop and 2-hop sharing the same
label have the highest homophily ratio.}
		\label{2hop1}
	\end{figure*}
	\section{Methodology}
	\subsection{Notation}
	Define graph data as $G=\lbrace \mathcal{V},\widetilde{A}, X\rbrace$, 
	where $\mathcal{V}$ represents the set of $N$ nodes, 
	$X=\lbrace x_1,...,x_N\rbrace^{\top}$ is the feature matrix. Initial graph structure $  \widetilde{A} $  
	characterizes the relation between nodes. $ D $ represents the degree matrix. 
	The normalized adjacency matrix is $  A  = D^{-\frac{1}{2}}(\widetilde{A} + I)D^{-\frac{1}{2}} $ and the corresponding graph Laplacian is $ L = I - A $. $1.$ is a matrix with all 1s.
	\subsection{Structure Reconstruction} \label{GC}
	
	In practice, we can't know whether a given graph is homophilic or not in unsupervised tasks. Hence, separately developing homophilic and heterophilic methods is unrealistic. Moreover, real graphs always contain both homophilic and heterophilic nodes. To have a holistic model, we develop a structure reconstruction             approach. Specifically, we construct a heterophilic graph and a homophilic graph from the original graph.

	\subsubsection{Heterophilic Graph Construction}
	Firstly, we select the nodes which are far away from each other in both feature space and structure space as negative pairs, which prevents us from false negative pairs. Specifically, we use complementary graphs of similarity graph and topology graph to construct a heterophily graph. The procedure is formulated as follows:
	\begin{equation}
		\begin{aligned}
			\bar{W} &= 1.-W,\\ 
			\bar{A} &=1.-A,\\ 
			H &=\bar{W} \odot \bar{A},
		\end{aligned}
	\end{equation}
	where the similarity matrix $W$ is obtained through cosine similarity of node features, which characterizes the closeness among nodes in feature space. $\odot$ represents the Hadamard product, which is used to describe non-neighbor relation in both feature space and topology space. 
    For homophilic graph, nodes of the same class tend to be adjacent topologically, and neighbors in complementary graph have a tendency to be different. The adjacent nodes in heterophilic graph are often dissimilar because of the connection between nodes of diverse types \cite{hop}. Moreover, nodes with large edge weights in similarity graph more likely belong to the same class
    , thus adjacent nodes in corresponding complementary graph are more possibly have small edge weights, i.e., they are different. It is rational to pick adjacent nodes from both $\bar{W}$ and
    $\bar{A}$, so that the reconstructed graph $H$ is heterophilic. $H$ could be dense, thus we just keep 5 edges for each node corresponding to top $5$ dissimilar nodes.
	\subsubsection{ Homophilic Graph Construction}
	In practice, even the homophilic graph doesn't have a homophily score of 1, i.e., there exists some heterophilic nodes in homophilic graph. Thus, we could further improve the homophily level of raw graph by minimizing the distances among adjacent nodes, which is formulated as:
	$$
    \min _{S_{i:}} \sum_{j=1}^N S_{i j}\left\|x_i-x_j\right\|^2,
    $$
    where $S_{i:}$ represents the $i$-th row of $S$. To avoid the trivial solution $S=I$, we rewrite above equation as: 
	$$
    \min _{ S_{i:}} \sum_{j=1}^NS_{i j}\left\|x_i-x_j\right\|^2+S_{i j}^2.
    $$
    It’s clear that the graph $S$ will be more homophilic when edges are defined by nodes sharing high similarity. The homophily ratio $h^{(l)}$ of nodes in different hops vary considerably according to Fig. \ref{2hop1}, where $h^{(l)}(G)=\frac{1}{|\tilde{A}_{ij}^{(l)}|}\sum_{\tilde{A}_{ij}^{(l)}>0}\mathbbm{1}(y_i=y_j)$ and $l$ is the number of hop and exponent of $\tilde{A}$  \cite{HM}. Consequently, the message propagation path in $S$ could be incorrect when the 2-hop neighbors of a node are dissimilar to its 1-hop neighbors. Furthermore, the nodes sharing the same label with 1-hop and 2-hop neighbors have the highest ratio. Based on this observation, we design a regularization term to integrate the 1-hop and 2-hop neighbor relation, i.e., we enforce all 2-hop neighboring  nodes are in the set of 1-hop neighborhood. Let $\|x_i - x_j\|^2=K_{ij}$, then we construct a homophilic graph $S$ by solving the following optimization problem:
    \begin{equation} \label{pr2}
    \begin{gathered}
    \min _{S_{ij}} S_{i j} K_{ij}+S_{i j}^2+\left(S_{i j}^{(2)}-S_{i j}\right)^2, \\
    \text { s.t. } S_{i j}>0, \sum_{j=1}^N S_{i j}=1,
    \end{gathered}
    \end{equation}
	where $S^{(2)}$ is the 2-hop graph, i.e., $S^{(2)}=S\times S$.
 \paragraph{Optimization}
	
	Firstly, we initialize $S$ with $A$. Then we reformulate problem (\ref{pr2}) via its Lagrangian function:
	\begin{equation} \label{hgl_lagrangian}
		\begin{aligned}
			\underset{S_{i:}}{\operatorname{min}}\sum_{j=1}^N[&S_{ij}K_{ij}+(S_{ij}^{(2)}-S_{ij})^2+S_{ij}^2]\\
			&-\sum _{j=1}^N\lambda_{1j}S_{ij}-\lambda_{2i}(\sum_{j=1}^NS_{ij}-1).
		\end{aligned}
	\end{equation}
	The derivative of Eq. (\ref{hgl_lagrangian}) w.r.t. $S_{ij}$ is 
	\begin{equation} \label{hgl_derivative}
		\begin{aligned}
			&K_{ij}+2(S_{ii}+S_{jj}-1)(S_{ij}^{(2)}-S_{ij})+2S_{ij}-\lambda_{1j}-\lambda_{2j}\\
			&+\sum_{f\neq j}^N2S_{jf}[S_{ij}S_{jf}+(S_{if}^{(2)}-S_{ij}S_{jf})-S_{if}].
		\end{aligned}
	\end{equation}
	Remove the self-loop on graph, i.e., let $S_{ii}=0$. $S$ and $S^{(2)}$ can be regarded as constants at the last iteration. By introducing $C_f$, 
	$$C_f=\begin{cases}S_{if}^{(2)}-S_{ij}S_{jf}-S_{if},i\neq f\\ 0,otherwise\end{cases}$$  
	 we rewrite Eq. (\ref{hgl_derivative}) as:
	\begin{equation} \label{hgl_derivative_simplify}
		K_{ij}-2(S_{ij}^{(2)}-S_{ij})+2S_{ij}-\lambda_{1j}-\lambda_{2i}+\sum_{f\neq j}2S_{jf}^2S_{ij}+2S_{jf}C_f.
	\end{equation}
	Note that the KKT condition 
 $\lambda_{1j}S_{ij}=0$ , which yields:
	\begin{equation} \label{hgl_derivative_simplify_2}
		\begin{aligned}
			S_{ij}&(K_{ij}-2(S_{ij}^{(2)}-S_{ij})+2S_{ij}-\lambda_{2j} \\&+\sum_{f\neq j}2S_{jf}^2S_{ij}+2S_{jf}C_f)=0.
		\end{aligned}
	\end{equation}
	Afterwards, we obtain the closed-form solution of $S_{ij}$:
	\begin{equation} \label{hgl_solution}
		S_{ij}=[\frac{2S_{ij}^{(2)}+\lambda_{2j}-K_{ij}-2\sum_{f \neq  j}S_{jf}C_f}{2(2+\sum_{f \neq j}S_{jf}^2)}]_+,
	\end{equation}
	where $[\bullet]_+$ operator means $\operatorname{max}(\bullet,0)$. 
 Following the second constraint, we have $$\sum_{j}[\frac{2S_{ij}^{(2)}+\lambda_{2j}-K_{ij}-2\sum_{f \neq j}S_{jf}C_f}{2(2+\sum_{f \neq j}S_{jf}^2)}]_+ -1 = 0.\\$$.

 \begin{figure*}[htbp]
		\centering
		\includegraphics[width=0.85\linewidth]{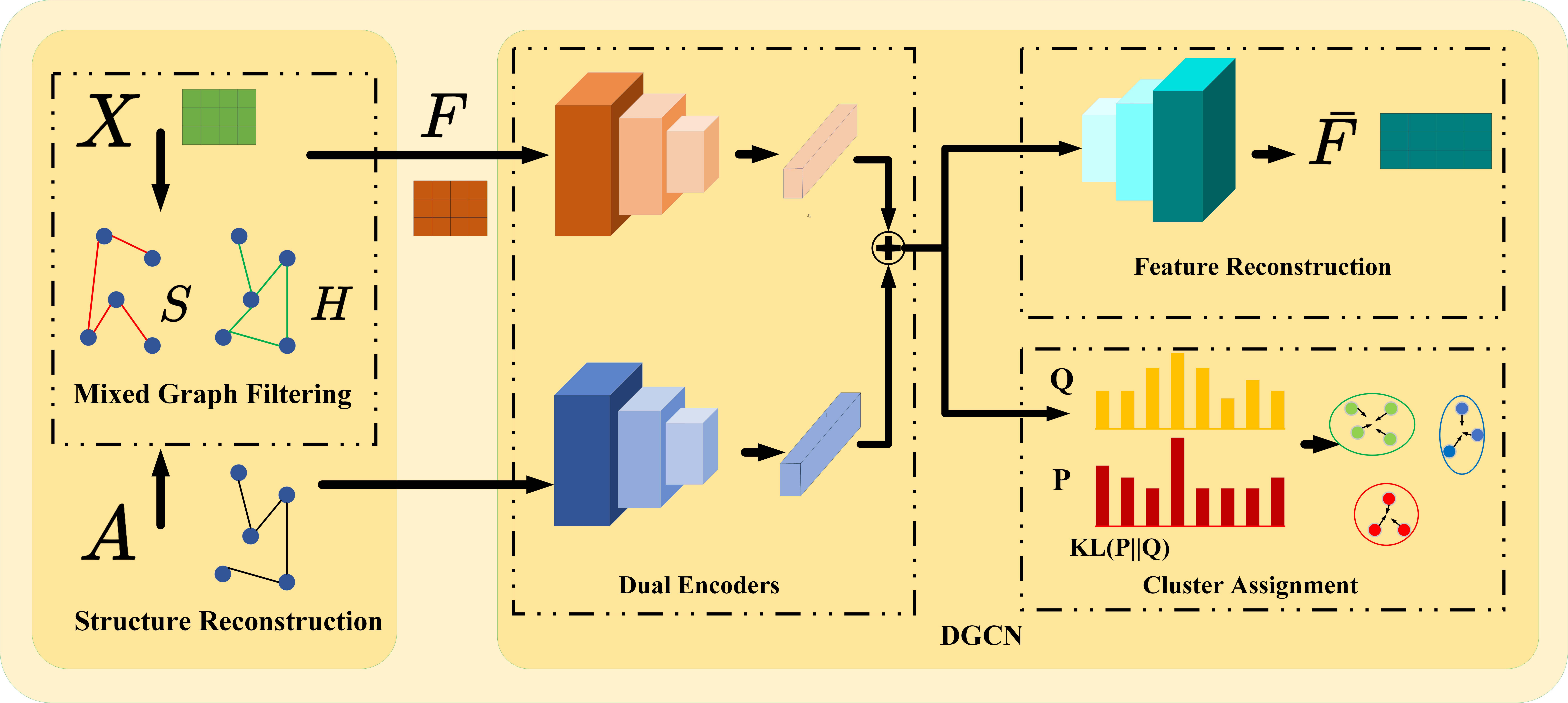}
		\caption{The framework of DGCN. There are three key components: 1) structure reconstruction is used to reconstruct homophilic and heterophilic graphs; 2) a mixed graph filter is designed to obtain smooth representations for any kind of graphs; 3) dual encoders are applied to learn embeddings in attribute and topology space.}
		\label{HGL}
	\end{figure*}		
	To obtain $S_{ij}$, we need to compute $\lambda_{2j}$ first by solving this optimization problem:
	\begin{equation} \label{hgl_lambda}
		\underset{\lambda_{2j}}{\operatorname{min}} \sum_{j=1}^N  [\frac{2S_{ij}^{(2)}+\lambda_{2j}-K_{ij}-2\sum_{f \neq j}S_{jf}C_f}{2(2+\sum_{f \neq j}S_{jf}^2)}]_+ -1.
	\end{equation}
	We can solve Eq. (\ref{hgl_lambda}) by gradient descent algorithm or treating it as a Linear programming (LP) problem. 
	\subsection{Graph Filtering}
 
	Based on the assumption that graph signal should be smooth, i.e., the neighbor nodes tend to be similar, low-pass filter has been used to obtain smoothed representations that are  clustering-friendly \cite{MCGC}. 
	One typical low-pass filter  \cite{AGC} can be formulated as:
	\begin{equation} \label{high_pass_filter}
		F = (I-\frac{1}{2}L)^k X,
	\end{equation}
	where $F$ is the filtered representation, $k$ is the order of graph filtering.\\
	However, Eq. (\ref{high_pass_filter}) could be ineffective resulting from its heavy dependence on raw topological graph which could be noisy and incomplete. Additionally, low-pass filtering neglects the high-frequency components in data, which leads to information loss and inferior performance. This would be more worse for heterophilic graph, where high-frequency information plays a critical role. Since it is impossible to know whether a given graph is homophilic or not in unsupervised learning, it's necessary to design a generic filter to handle different types of graphs. To this end, we design a mixed filter for graph data as follows:
	\begin{equation} \label{combine_filter}
		F = \mu (\frac{1}{2}L_H)^k X + (1-\mu) (I - \frac{1}{2}L_S)^{k} X,
	\end{equation}
	where $\mu>0$ is a trade-off parameter balancing low-pass and high-pass representations, $L_S$ and $L_H$ are the normalized Laplacian matrices of reconstructed homophilic and heterophilic graphs. Note that our mixed filter is not combining low-pass and high-pass filter simply, we apply newly constructed graphs rather than the raw graph, which often has low-quality. The final representation $F$ is used as the input of  clustering network.\\
	 \subsection{Dual Graph Clustering Network}
		In this section, we introduce our proposed Dual Graph Clustering Network (DGCN) to address the nodes clustering task. As shown in Fig. \ref{HGL}, DGCN contains two unshared encoders and a decoder, and all of them are MLPs. Different from GNNs which learn representations in only one space, two encoders $E_{\theta_F}$ and $E_{\theta_A}$  are utilized to map filtered features and structural graph into attribute subspace and structure subspace respectively. This would reduce the interaction between attribute and structure for heterophilic nodes. The structure encoder is applied to preserve some original structure information. The obtained representations are:
	\begin{equation}
		\begin{aligned}
			Z_F = E_{\theta_F}(F),\\
			Z_A = E_{\theta_A}(A).
		\end{aligned}
	\end{equation}
 Moreover, to alleviate representation collapse, i.e., representations of all nodes tend to be the same, we add a correlation reduction item to prevent it \cite{BarlowTwins}.

	\begin{equation}
		\begin{aligned}
			\mathcal{L}_{CR} &=\frac{1}{d^2} \sum\left(\mathbf{M}-\widetilde{\mathbf{I}}\right)^2 \\
			&=\frac{1}{d^2} \sum_{i=1}^d\left(\mathbf{M}_{i i}-1\right)^2+\frac{1}{d^2-d} \sum_{i=1}^d \sum_{j \neq i}\mathbf{M}_{i j}^2,
		\end{aligned}
	\end{equation}
	where $d$ is the dimension of node attribute, $M$ is the similarity of corresponding nodes in two encoders, $$M_{ij}=\frac{Z_{A}{}_{i}^{\top}Z_F{}_{j}}{\|Z_{A}{}_{i}\|\cdot\|Z_F{}_{j}\|}.$$ Afterwards, $Z_A$ and $Z_F$ are concatenated as $Z$.

 \begin{table*}[t]
		\centering
	\caption{Statistics information of datasets.}
		\label{tab:addlabel}%
    
    \begin{tabular}{ccccccc}
    \toprule
    \multicolumn{2}{c}{Graph datasets} & Nodes & Dims. & Edges & Clusters & Homophily Ratio \\
    \midrule
    {Heterophilic Graphs} & Texas & 183   & 1703  & 325   & 5& 0.0614  \\
          & Cornell & 183   & 1703  & 298   & 5     & 0.1220  \\
          & Wisconsin & 251   & 1703  & 515   & 5      & 0.1703  \\
          & Washington  & 230   & 1703  & 786   & 5    & 0.1434 \\
          & Twitch & 1912 & 2545 & 31299 & 2    & 0.5660 \\
          & Squirrel & 5201  & 2089  & 217073 & 5    & 0.2234  \\
    \midrule
    {Homophilic Graphs} & Cora  & 2708  & 1433  & 5429  & 7    & 0.8137  \\
          & Citeseer & 3327  & 3703  & 4732  & 6     & 0.7392  \\
          & ACM   & 3025  & 1870  & 29281 & 3     & 0.8207 \\
          & AMAP  & 7650  & 745   & 119081 & 8    & 0.8272 \\
          & EAT   & 399   & 203   & 5994  & 4     & 0.4046 \\
    \bottomrule
    \end{tabular}%
		
	\end{table*}
 
	Decoder is employed to obtain reconstructed features $\bar{F}$. The features of some ``easy samples" change little during reconstruction, which indicates that these nodes contribute less information for training our model. We adopt the Scaled Cosine Error (SCE) as the objective of reconstruction \cite{GraphMAE}, which can down-weight easy samples’ contribution by controlling sharpening parameter $\beta$   in training:
	\begin{equation}
		\mathcal{L}_{\mathrm{SCE}}=\sum^N_{i=1 }\left(1-\frac{F_i^\top \bar{F}_i}{\left\|F_i\right\| \cdot\left\|\bar{F}_i\right\|}\right)^\beta,  
 \beta \geq 1.
	\end{equation}
	Finally, we pull soft cluster assignment probabilities distribution and target distribution for cluster enhancement. Specifically, the soft assignment distribution $Q$ is computed as:
	\begin{equation}
		q_{i u}=\frac{\left(1+\left\|z_i-\sigma_u\right\|^2 / \alpha\right)^{-\frac{\alpha+1}{2}}}{\sum_{u^{\prime}}\left(1+\left\|z_i-\sigma_{u^{\prime}}\right\|^2 / \alpha\right)^{-\frac{\alpha+1}{2}}},
	\end{equation}
	where cluster centres $\sigma$ are initialized by $k$-means on embeddings and $\alpha$ is the Student’s $t$-distribution’s degree of freedom. Then target distribution $P$ is formulated as:
	\begin{equation}
		p_{i u}=\frac{q_{i u}^2 / \sum_i q_{i u}}{\sum_{k'}\left(q_{i k'}^2 / \sum_i q_{i k'}\right)}.
	\end{equation}
	
	We minimize the KL divergence loss
	between $Q$ and $P$ distributions to make the data representation closer to cluster centres
	and improve cluster cohesion:
	\begin{equation}
		\mathcal{L}_{CLU}=K L(P \| Q)=\sum_i \sum_u p_{i u} \log \frac{p_{i u}}{q_{i u}}
	\end{equation}
	In summary, the objective of DGCN can be computed by:
	\begin{equation}
		\mathcal{L} = \mathcal{L}_{CR} + \mathcal{L}_{SCE} + \mathcal{L}_{CLU}.
	\end{equation}
	We minimize this objective function to train our model and the result of clustering for node $i$ is:
	\begin{equation}
		Y_{i} = \underset{c}{\operatorname{argmax}q_{ic}}.
	\end{equation}


	\section{Experiments}
	
	\subsection{Datasets}

	To evaluate the effectiveness of the proposed method, we conduct extensive experiments on 11 benchmarks, including homophilic graph datasets, like Cora, Citeseer \cite{GAE}, ACM \cite{O2MAC}, 
 AMAP \cite{DCRN}, EAT
 \cite{UATEatBat}; heterophilic graph datasets, like Texas, Cornell, Wisconsin, Washington  \cite{Geom-GCN}, Twitch \cite{lim1}, and Squirel \cite{squirel}. The statistical information of them is summarized in Table 1. 
	

	 \begin{table*}[htbp]
		\centering
  \caption{Results on heterophilic graphs. The best results are \textbf{bolded} with \textcolor{red}{red} and the second-best performance is also \textbf{bolded}. `-' means that the source code can't produce any results.}
		\label{Re2}%
		 
    \begin{tabular}{ccccccccccccc}
    \toprule
    \multirow{2}[4]{*}{Methods} & \multicolumn{2}{c}{Texas} & \multicolumn{2}{c}{Cornell} & \multicolumn{2}{c}{Wisconsin} & \multicolumn{2}{c}{Washington} & \multicolumn{2}{c}{Twitch} & \multicolumn{2}{c}{Squirrel} \\
\cmidrule{2-13}          & ACC   & NMI   & ACC   & NMI   & ACC   & NMI   & ACC   & NMI   & ACC   & NMI   & ACC   & NMI \\
    \midrule
    DAEGC & 45.99  & 11.25 & 42.56  & 12.37 & 39.62  & 12.02 & 40.43  & -   & 56.59  & -   & 25.55  & 2.36 \\
    MSGA  & 57.22  & 12.13 & 50.77  & 14.05 & 54.72  & 16.28 & -   & -   & -   & -   & 27.42  & 4.31 \\
    FGC   & 53.48  & 5.16  & 44.10  & 8.6   & 50.19  & 12.92 & 51.30  & -   & \textbf{70.71 } & -   & 25.11  & 1.32 \\
    GMM & 58.29 & 13.06 & 58.86 & -   & 52.08 & 8.89  & 60.86  & 20.56 & -   & -   & -   & - \\
    RWR & 57.22 & 13.87 & 58.29 & -   & 53.96 & 16.02 & 63.91  & 23.13 & -   & -   & -   & - \\
    ARVGA & 59.89  & 16.37 & 56.23  & -   & 56.23  & 13.73 & 60.87  & 16.19 & -   & -   & -   & - \\
    
    DGCN$_{\beta=2}$ & \textbf{72.68 } & \textbf{33.67} & \textcolor[rgb]{ 1,  0,  0}{\textbf{62.29 }} & \textcolor[rgb]{ 1,  0,  0}{\textbf{29.93}} & \textbf{71.71 } & \textbf{41.29} & \textbf{69.13 } & \textbf{28.22} & 70.34  & \textbf{39.84} & \textbf{31.34 } & \textbf{7.24} \\
    DGCN$_{\beta=1}$ & \textcolor[rgb]{ 1,  0,  0}{\textbf{74.57 }} & \textcolor[rgb]{ 1,  0,  0}{\textbf{39.93}} & \textbf{61.74 } & \textbf{21.76} & \textcolor[rgb]{ 1,  0,  0}{\textbf{72.90 }} & \textcolor[rgb]{ 1,  0,  0}{\textbf{43.52}} & \textcolor[rgb]{ 1,  0,  0}{\textbf{69.56 }} & \textcolor[rgb]{ 1,  0,  0}{\textbf{28.43}} & \textcolor[rgb]{ 1,  0,  0}{\textbf{71.00 }} & \textcolor[rgb]{ 1,  0,  0}{\textbf{41.81}} & \textcolor[rgb]{ 1,  0,  0}{\textbf{31.39 }} & \textcolor[rgb]{ 1,  0,  0}{\textbf{8.54}} \\
    \bottomrule
    \end{tabular}%
	\end{table*}%
	\subsection{ Comparison Methods}
	To demonstrate the superiority of our method, we adopt 13 baselines for performance comparison. Specifically, there are four kinds of methods: 1) typical GNN-based methods, like DAEGC \cite{DAEGC}, MSGA \cite{MSGA}, SSGC \cite{S2GC}, GMM \cite{CDRS}, RWR \cite{RWR}, ARVGA \cite{ARVGA}; 2) contrastive learning-based methods, like MVGRL \cite{MVGRL}, SDCN \cite{SDCN}, DFCN \cite{DFCN}, and DCRN \cite{DCRN}, which employ MLP and GNNs jointly to learn an aligned representation from augmented views; 3) state-of-the-art MLP-based clustering method AGE \cite{AGE}, which obtains a clustering-favorable representation via Laplacian smoothing filter and adaptive encoder; 4) shallow methods which utilize a low-pass filter to smooth the raw features and remove the noises, like MCGC \cite{MCGC} and FGC \cite{FGC}. We implement DGCN with both $\beta=1$ and $\beta=2$. In fact, SCE loss with $\beta=1$ is equivalent to traditional Frobenius norm. \\
 \subsection{Experimental Setting}
	For fairness, all compared methods are implemented with the same setting in original papers. Our network is trained with Adam optimizer for 500 epochs until convergence. The learning rate of optimizer is set to 1e-2. We tune filter order $k$ in [1, 2, 3, 4, 5, 10]. 
  We adopt ACCuracy (ACC) and Normalized Mutual Information (NMI) as cluster metrics in all experiments.
 \subsection{ Results}	
 
\begin{table*}[!htbp]
		\centering
  \caption{Results on homophilic graphs. AvgRank represents the average ranking of methods on five graphs.}
		\label{Re1}%
        \begin{tabular}{cccccccccccc}
    \toprule
    \multirow{2}[4]{*}{Methods} & \multicolumn{2}{c}{Cora} & \multicolumn{2}{c}{Citeseer} & \multicolumn{2}{c}{ACM} & \multicolumn{2}{c}{AMAP} & \multicolumn{2}{c}{EAT} & \multirow{2}[4]{*}{AvgRank} \\
\cmidrule{2-11}          & ACC   & NMI   & ACC   & NMI   & ACC   & NMI   & {ACC} & NMI   & ACC   & NMI   &  \\
    \midrule
    DFCN  & 36.33  & 19.36 & 69.50  & 43.9  & 90.90  & 69.40  & {\textbf{76.88}} & \textbf{69.21} & 49.37 & \textcolor[rgb]{ 1,  0,  0}{\textbf{32.90 }} & 5.4 \\
    DCRN  & 48.93  & -   & 70.86  & \textcolor[rgb]{ 1,  0,  0}{\textbf{45.86}} & 91.93  & 71.56 & {\textcolor[rgb]{ 1,  0,  0}{\textbf{79.94}}} & \textcolor[rgb]{ 1,  0,  0}{\textbf{73.70 }} & 51.33 & -   & 3.6 \\
    SSGC  & 69.60  & 54.71 & 69.11  & 42.87 & 89.09  & 64.71 & {60.23} & 60.37 & 32.41 & 4.65  & 7.6 \\
    MVGRL & 70.47  & 55.57 & 68.66  & 43.66 & 86.73  & 60.87 & {45.19} & 36.89 & 32.88 & 11.72 & 8.2 \\
    SDCN  & 60.24  & 50.04 & 65.96  & 38.71 & 90.45  & 68.31 & {53.44} & 44.85 & 39.07 & 8.83  & 7.8 \\
    AGE   & \textcolor[rgb]{ 1,  0,  0}{\textbf{73.50 }} & \textcolor[rgb]{ 1,  0,  0}{\textbf{57.58}} & 70.39  & 44.92 & 90.91  & 69.42 & {75.98} & -   & 47.26 & -   & 4 \\
    MCGC  & 42.85  & 24.11 & 64.76  & 39.11 & 91.64  & 70.71 & {71.64} & 61.54 & 32.58 & 7.04  & 7.6 \\
    FGC   & \textbf{72.90 } & 56.12 & 69.01  & 44.02 & 88.13  & 62.77 & {71.04} & -   & 41.11 & -   & 6.2 \\
    DGCN$_{\beta=2}$ & 72.19 & 56.04 & \textbf{71.27 } & 44.13 & \textbf{92.03 } & \textbf{71.58} & 76.07 & 66.13 & \textbf{53.13} & 22.92 & \textbf{2.6} \\
       DGCN$_{\beta=1}$ & 72.89  & \textbf{56.82} & \textcolor[rgb]{ 1,  0,  0}{\textbf{71.60 }} & \textbf{44.83} & \textcolor[rgb]{ 1,  0,  0}{\textbf{92.60 }} & \textcolor[rgb]{ 1,  0,  0}{\textbf{71.85}} & {76.06} & 65.46 & \textcolor[rgb]{ 1,  0,  0}{\textbf{53.52}} & \textbf{24.81} &\textcolor[rgb]{ 1,  0,  0}{\textbf{ 2}} \\
    \bottomrule
    \end{tabular}%
		
	\end{table*}%

\begin{figure*}[t]
		\centering
		\subfigure[Cora with mixed filter]{
			\includegraphics[width=0.235\linewidth]{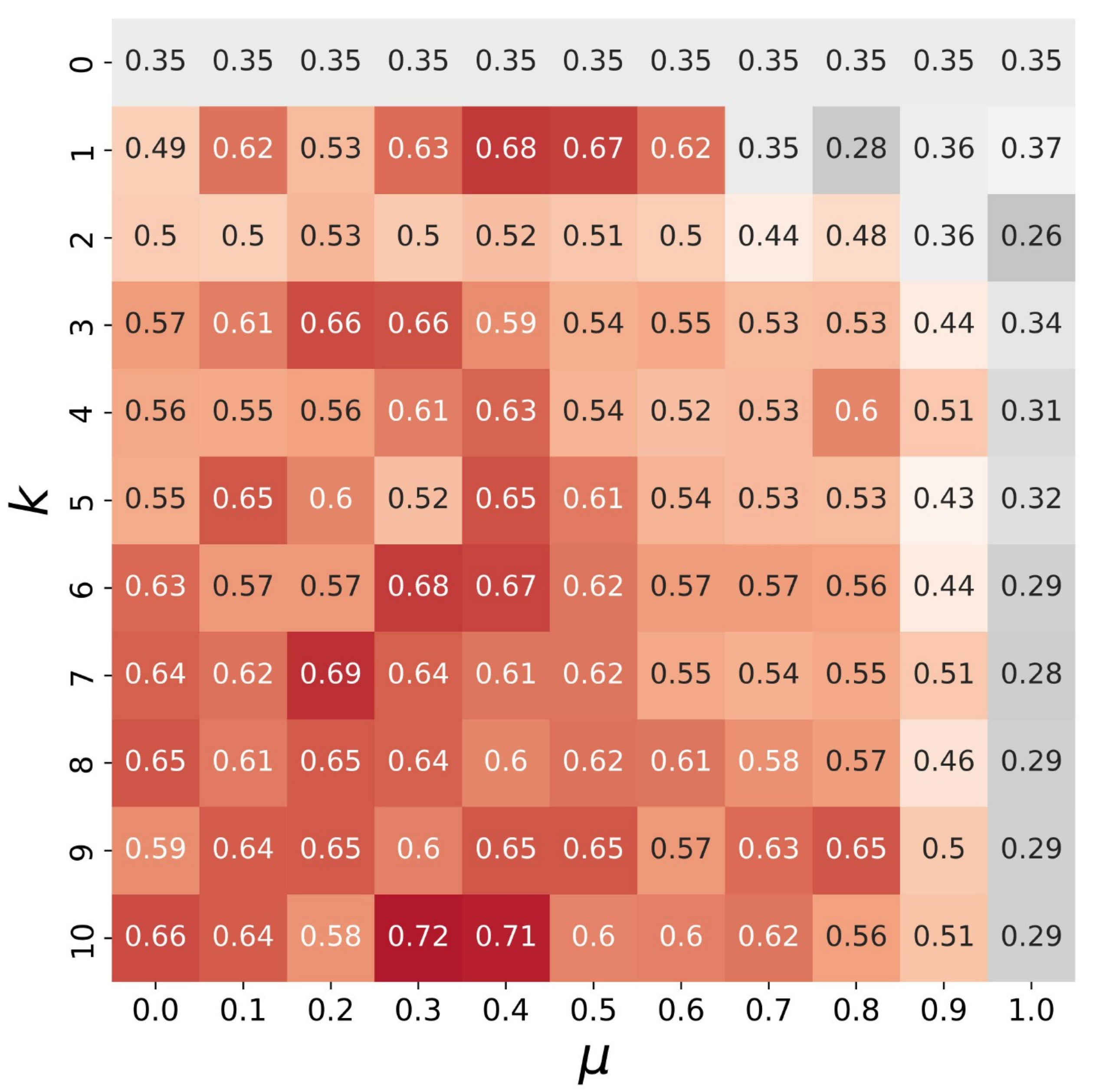}
		}
		\subfigure[Cora   with raw graph $A$]{
			\includegraphics[width=0.235\linewidth]{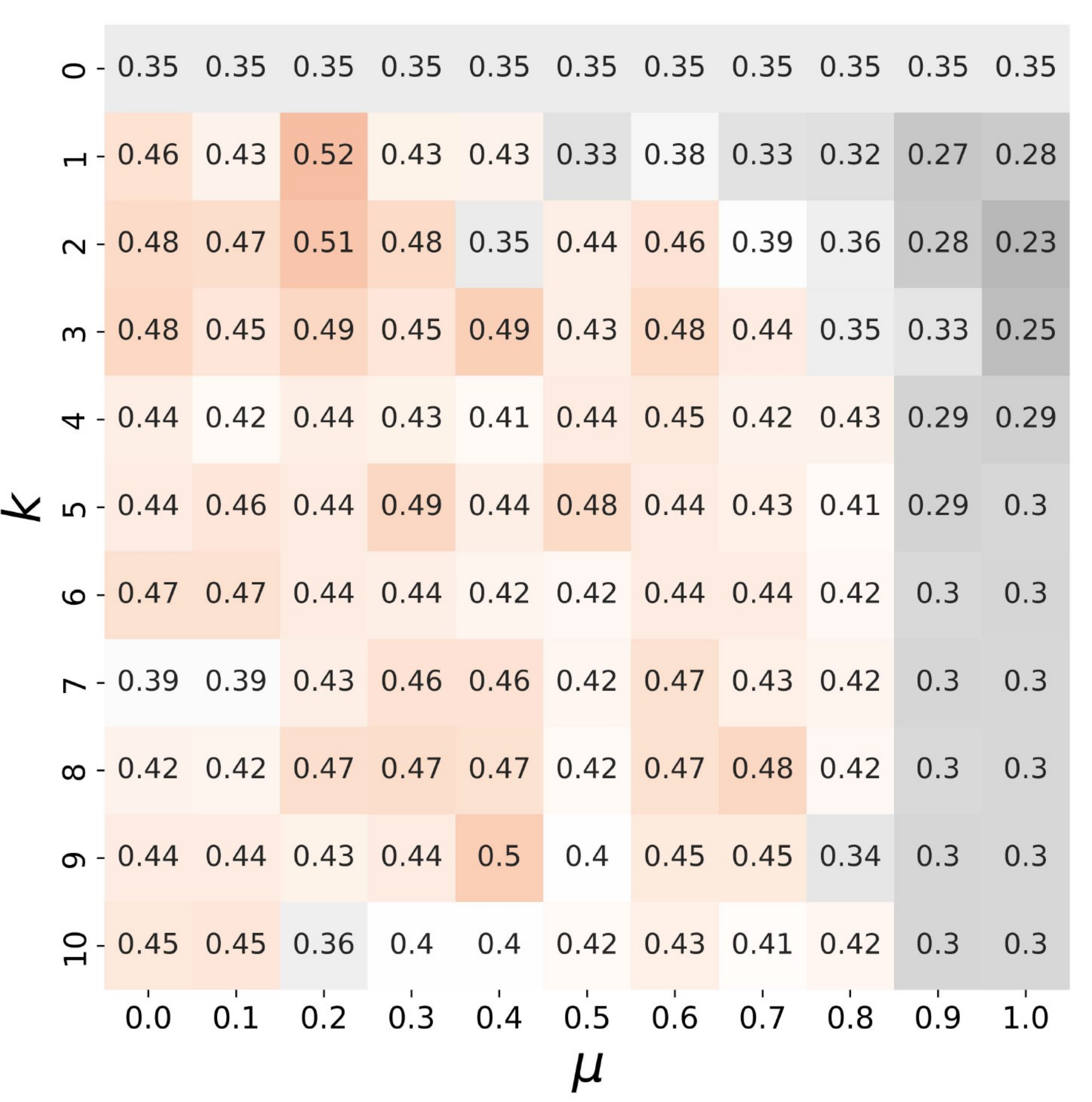}
		}
		\subfigure[Texas  with mixed filter]{
			\includegraphics[width=0.235\linewidth]{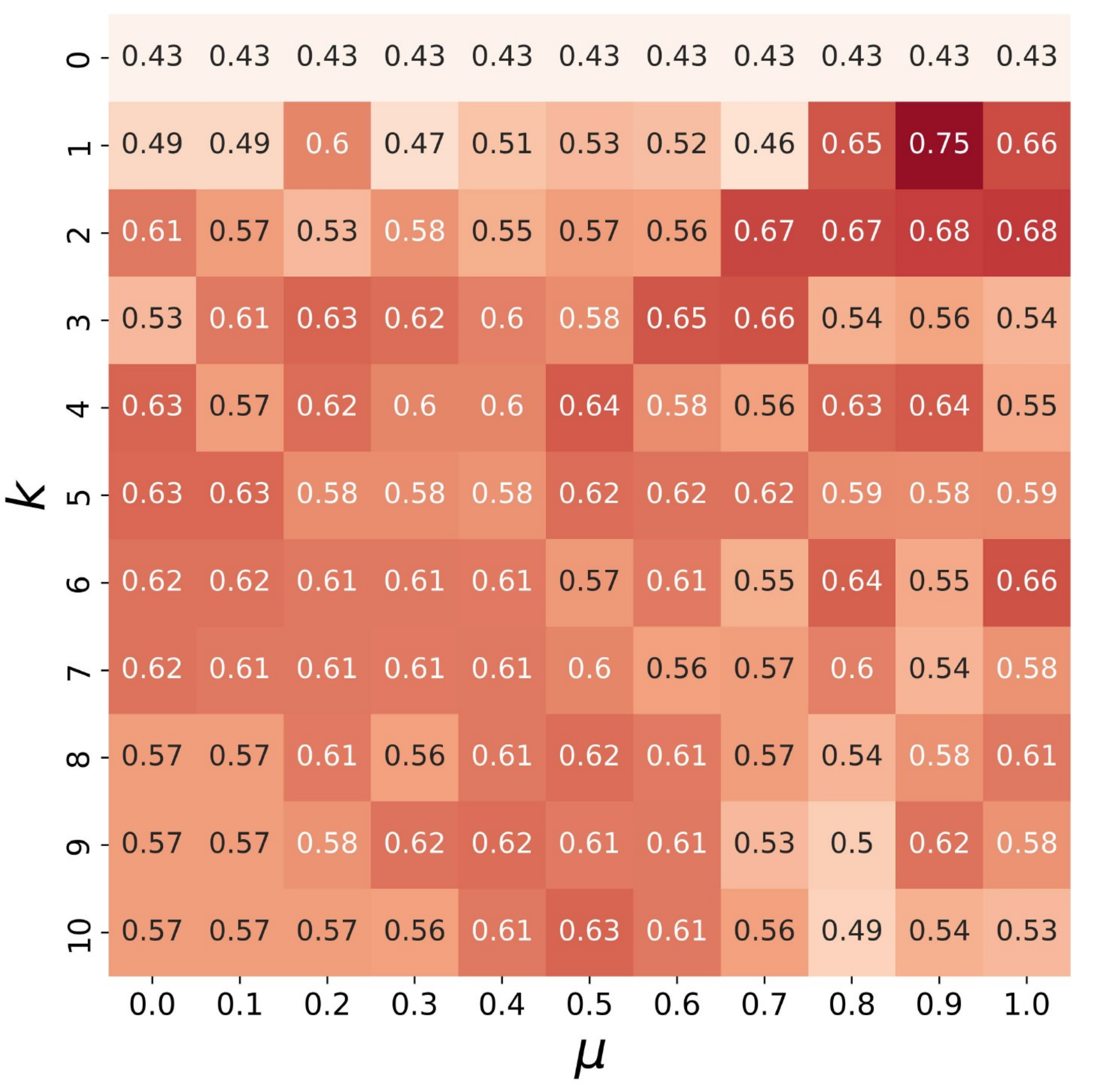}
		}
		\subfigure[Texas  with raw graph $A$]{
			\includegraphics[width=0.235\linewidth]{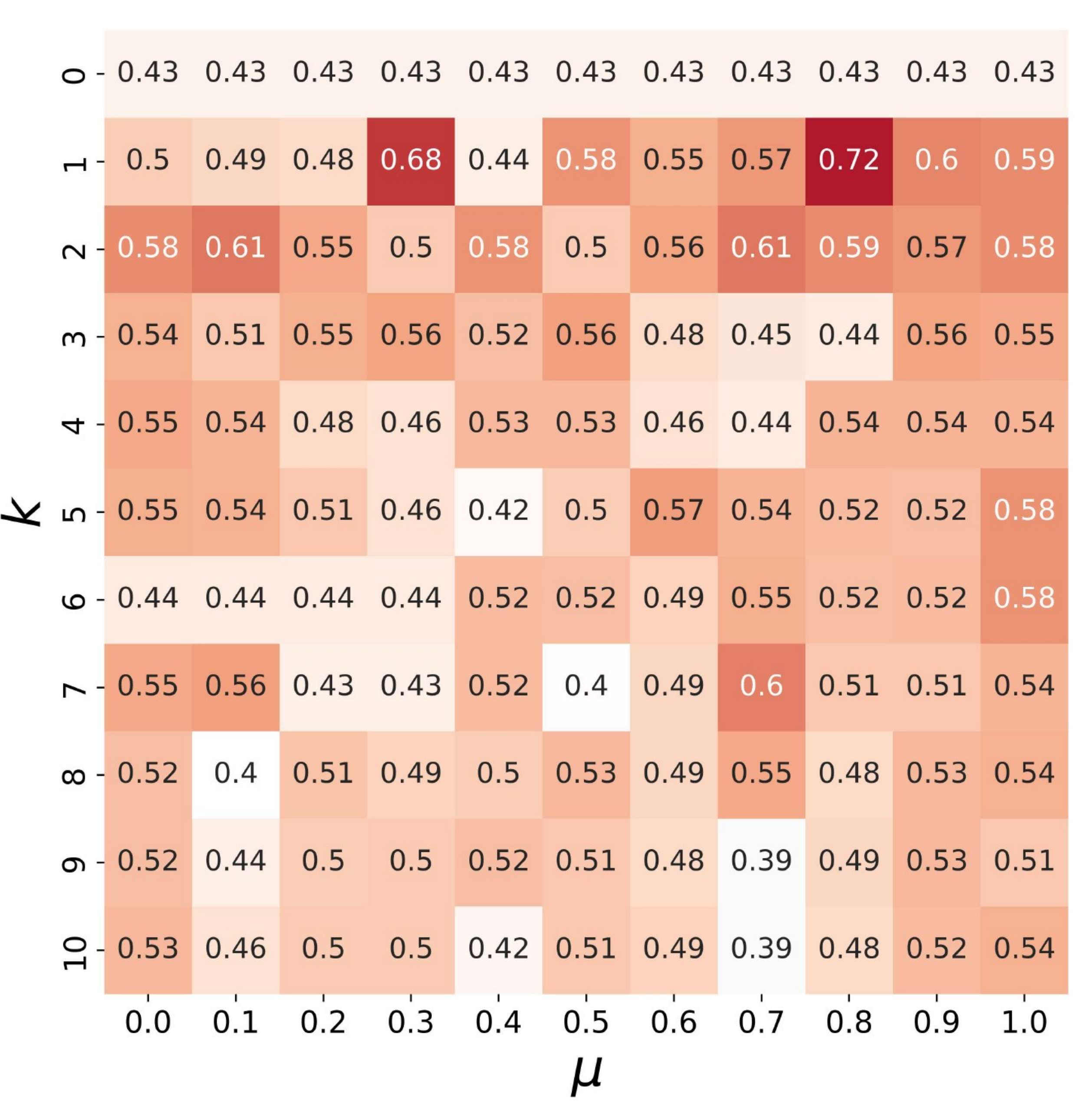}
		}
		\caption{Accuracy on Cora and Texas with raw graphs and reconstructed graphs.}
		\label{Re3}
	\end{figure*}
 Table 2 reports the results on heterophilic datasets. We can see that DGCN obtains much better performance than existing clustering methods. With respect to the second-best performance, our method improves the accuracy up to $10\%$ on Texas and Wisconsin, $5\%$ on Washington. One reason is that the comparison methods are not deliberately designed for heterophilic graphs. To our best knowledge, there is no clustering techniques considering graph heterophily in the literature. In addition, GNN-based methods don’t perform well on heterophilic graphs, which is consistent with the conclusion on classification task. By contrast, our method takes care of both homophilic and heterophilic nodes, thus it improves the performance significantly. In addition, we can see that $\beta$ can impact the performance a little bit.
 

 Table 3 shows the clustering results on homophilic graphs. Our method achieves competitive performance and shows the highest rank. 
 It can be observed that the SOTA contrastive learning methods produce unstable results. This is because their performance highly depends on the graph augmentation strategy, which requires domain knowledge and is not flexible to any data. For shallow methods FGC and MCGC, their performance also changes a lot on different datasets. This is attributed to their usage of low-pass filter, which is not suitable to real graphs with different levels of homophily. DGCN is better than AGE in most cases since our mixed filter explores both low-frequency and high-frequency components of original data and reduces information loss, while AGE neglects the high-frequency signal. 
 
 To sum up, DGCN obtains stable and promising results on both homophilic and heterophilic graphs. This is mainly because it extracts homophilic and heterophilic information from raw graph. Consequently, DGCN is applicable to cluster real graphs, where we have no idea of homophily ratios.




	 \section{Ablation Study}
 \begin{figure*}[htbp]
		\centering
		\subfigure[$A$ of Cora]{
			\includegraphics[width=0.23\linewidth]{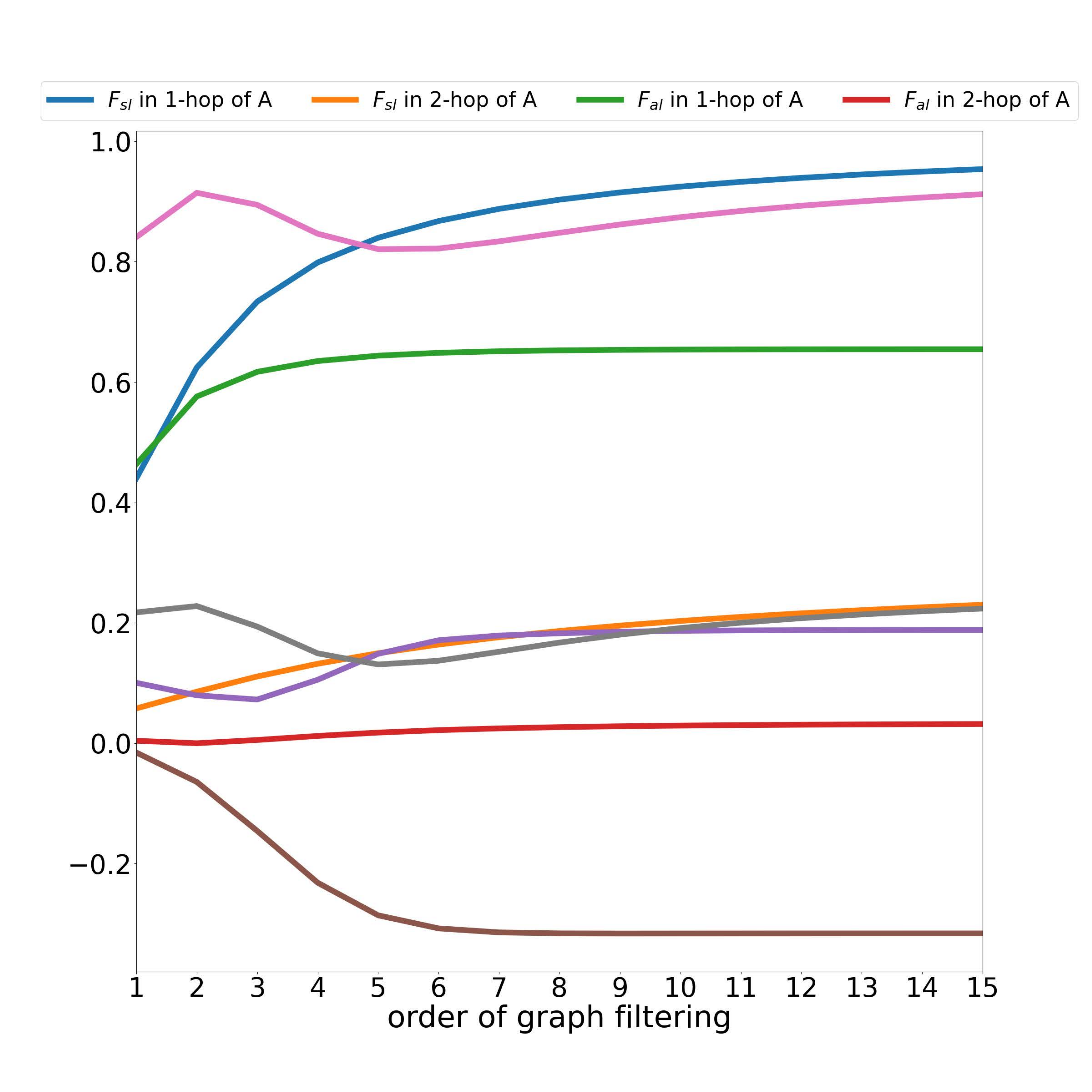}
		}
            \subfigure[$A$ of Texas]{
			\includegraphics[width=0.23\linewidth]{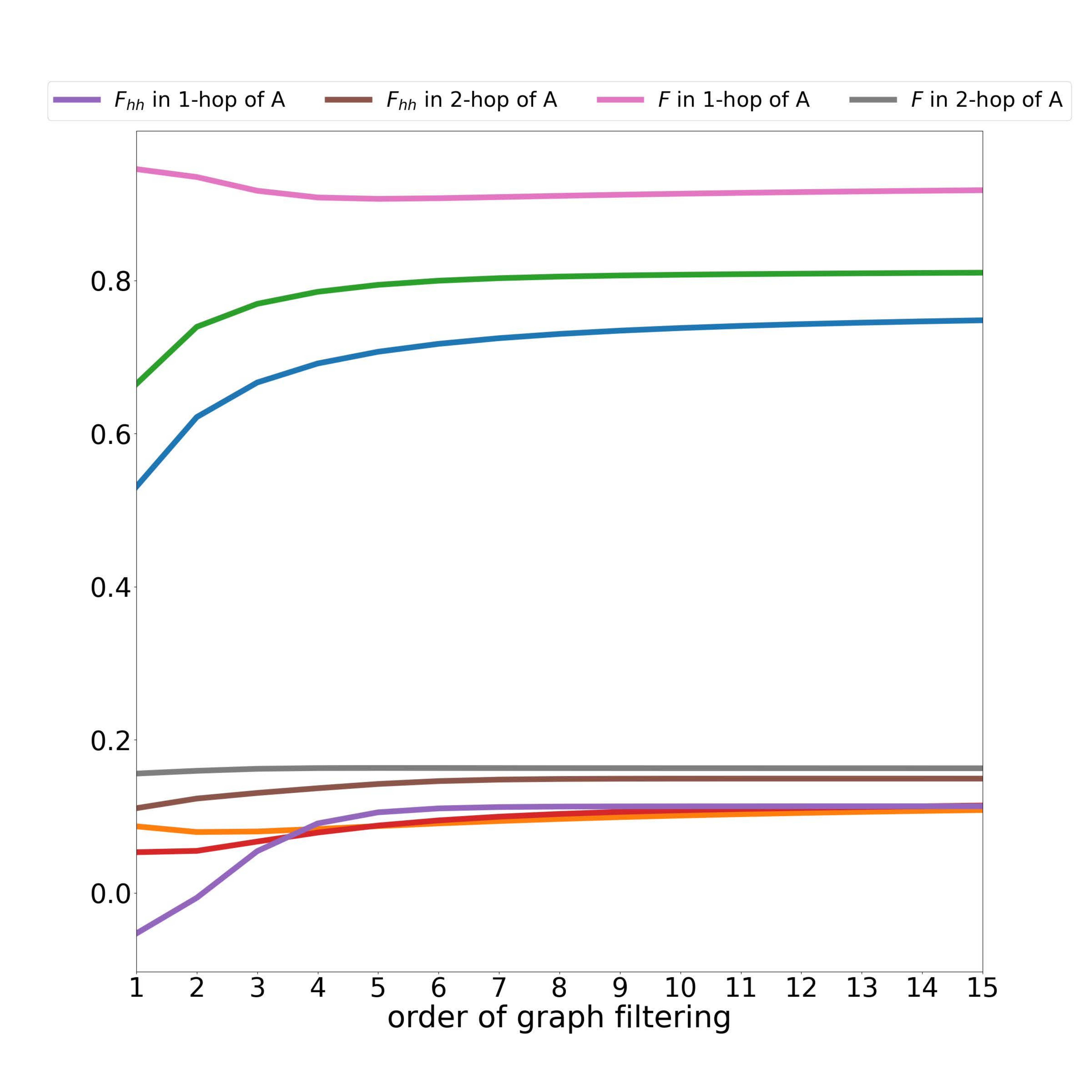}
		}
              \subfigure[$S$ of Cora]{
            			\includegraphics[width=0.23\linewidth]{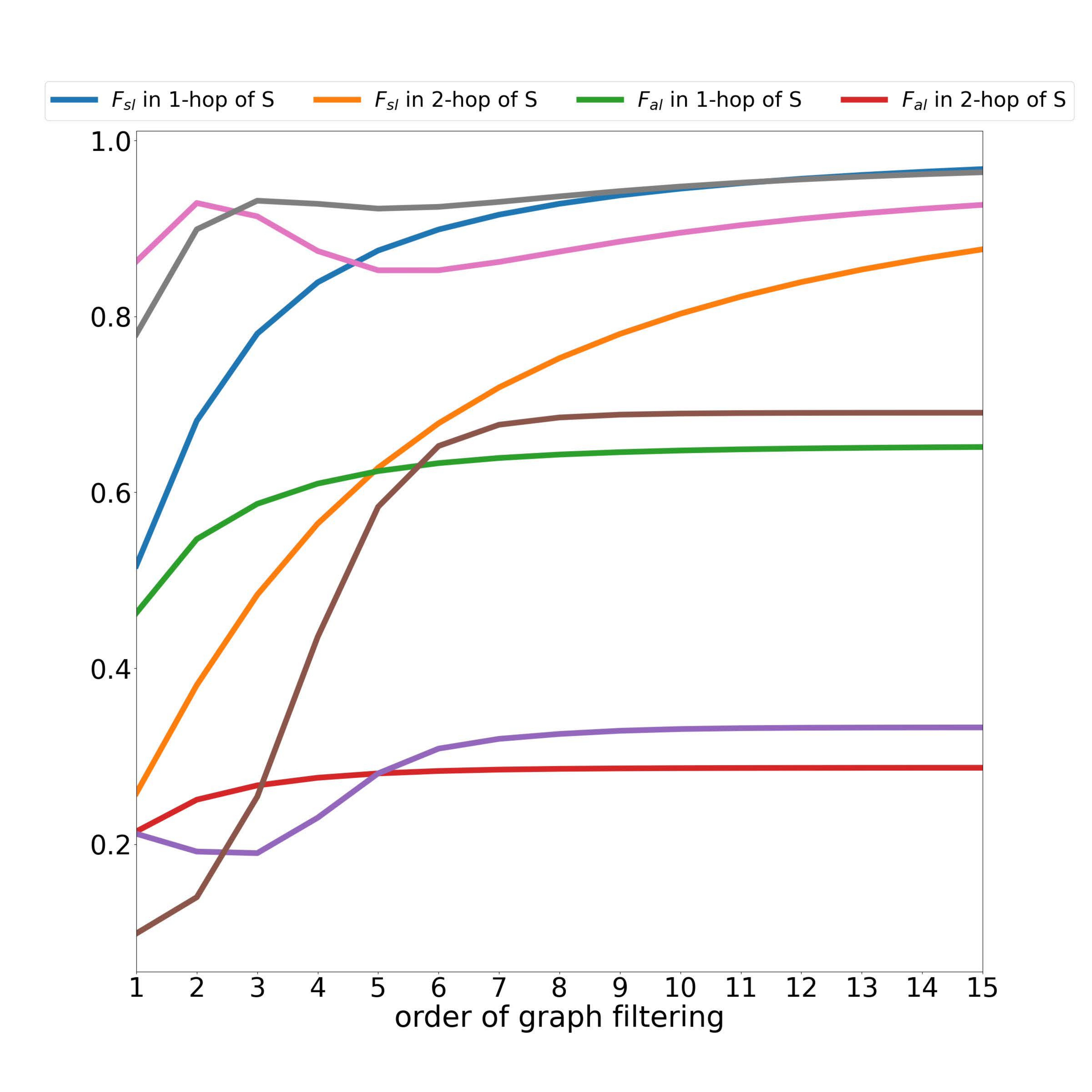}
            		}
		\subfigure[$S$ of Texas]{
			\includegraphics[width=0.23\linewidth]{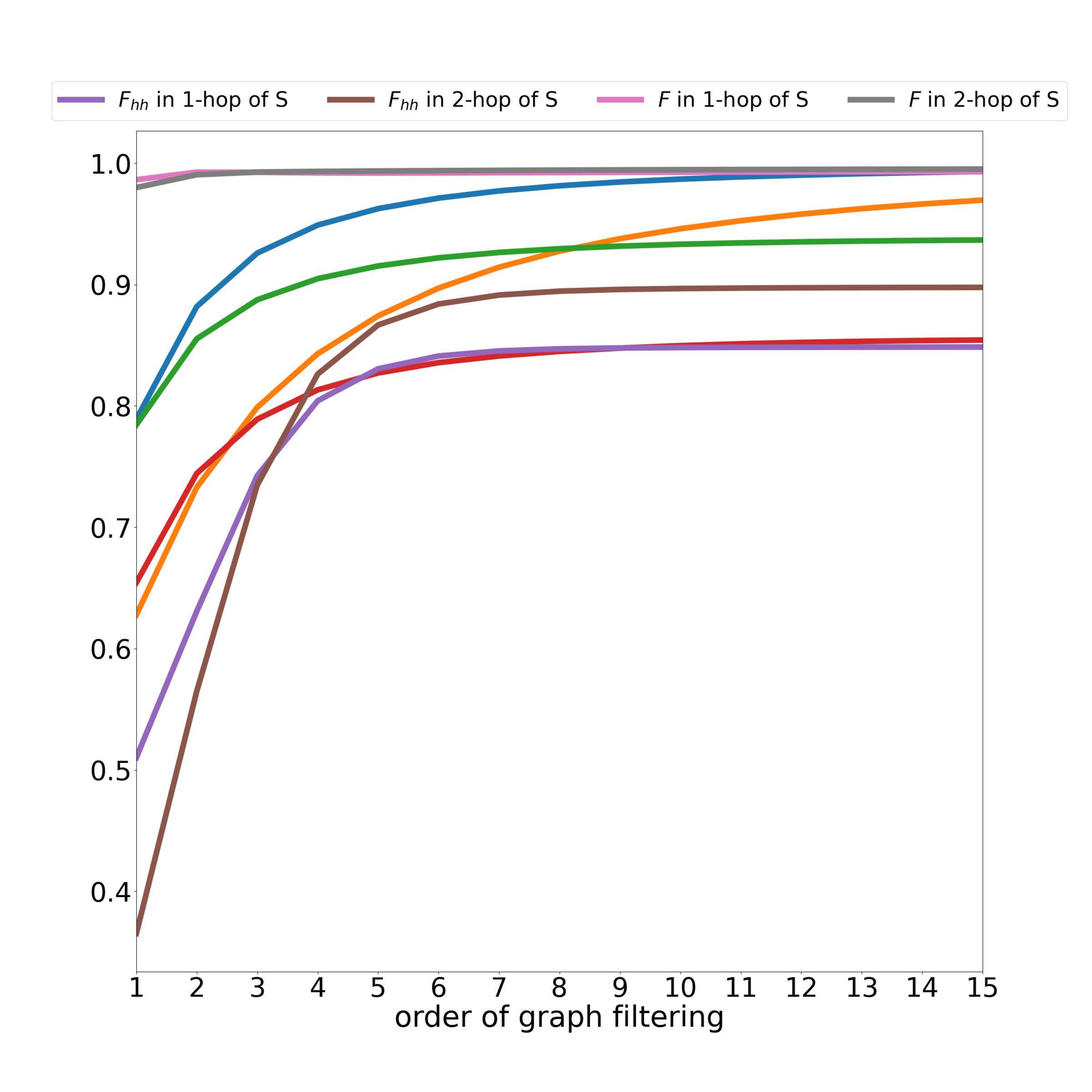}
		}
		\caption{The variation of average similarity of 1-hop and 2-hop neighbor nodes under different filter orders on Cora and Texas. $F_{sl}$ and $F_{al}$ represent the filtered features with low-pass filter on homophilic graph $S$ and raw graph $A$, $F_{hh}$ is obtained by high-pass filter on heterophilic graph $H$. 
   }
  
		\label{GF_SA}
	\end{figure*}

	Our mixed filter is based on the newly constructed homophilic and heterophilic graphs in which neighbor nodes tend to be similar and dissimilar, thus they contain low-frequency and high-frequency information, respectively. To show the effectiveness of our filter, we examine the clustering accuracy of DGCN with different filter orders $k$ and balance parameters $\mu$. Particularly, we don’t apply graph filtering when $k=0$ and employ low-pass (high-pass) filter only when $\mu=0$ ($\mu=1$). As shown in Fig. \ref{Re3}, graph filtering considerably boosts model performance on both homophilic and heterophilic graphs with $A$ or reconstructed graphs. Furthermore, the best performance is always achieved when $\mu\neq0$, which validates the importance of incorporating high-frequency information. 
 
 With respect to raw graph, the constructed graphs further boost the performance. However, there are two differences: 1) heterophilic graph is sensitive to high-frequency components while homophilic graph prefers to low-frequency parts; 2) DGCN achieves the best performance on heterophilic graph with a small $k$, while it also works well with a large $k$ on homophilic graph.\\

\begin{table}[htbp]
		\centering
		\footnotesize
		\caption{The  homophily score of constructed graphs.}
		\setlength{\tabcolsep}{1mm}{\begin{tabular}{cccc}
    \toprule
          & A & S & H \\
    \midrule
    Cora  & 0.8137  & 0.8110  & 0.1699  \\
    Citeseer & 0.7392  & 0.8018$\uparrow$  & 0.1780  \\
    ACM   & 0.8207  & 0.9006$\uparrow$  & 0.3236  \\
    AMAP  & 0.8272 & 0.9449$\uparrow$ & 0.1259 \\
    EAT   & 0.4046 & 0.6039$\uparrow$ & 0.2691 \\
    \midrule
    Texas & 0.0614  & 0.4654$\uparrow$  & 0.1767  \\
    Cornell & 0.1220  & 0.4583$\uparrow$  & 0.1734  \\
    Wisconsin & 0.1703  & 0.4301$\uparrow$  & 0.2274  \\
    Washington  & 0.1434 & 0.4522${\uparrow}$ & 0.1391$\textcolor{green}{\downarrow}$\\
    Twitch & 0.5660  & 0.7103$\uparrow$ & 0.2401$\textcolor{green}{\downarrow}$\\
    Squirrel & 0.2234  & 0.4781$\uparrow$  & 0.1999$\textcolor{green}{\downarrow}$\\
    \bottomrule
    \end{tabular}}
	\end{table}%
 
We report the homophily score of constructed graphs in Table. 4. Homophilic graph $S$ indeed has higher homophily score on most datasets, which proves that the designed regularizer pulls 1-hop and 2-hop neighbors close. The homophily score of heterophilic graph is improved by 16\% on average. There is no clear improvement on Cora because few nodes are near each other in terms of a 1-hop and 2-hop manner. All $H$s have a relative small value. However, the homophily score of $H$ is increased on Texas, Cornell, and Wisconsin. The reason is that these graphs have few edges, then reconstructed  heterophilic graphs become dense, where some edges connect the nodes of the same class. 
 
    To further understand the influence of filters, we plot the variation of cosine similarity by changing the filter order $k$ in Fig. \ref{GF_SA}. 
    As observed, on Cora, filtered feature similarity of 1-hop and 2-hop nodes obtained from high-order low-pass filter gets big while features from high-order high-pass filter tend to be dissimilar. By contrast, on heterophilic graph Texas, features from high-order low-pass filter or high-pass filter get more similar. These verify that graph filtering does preserve different kinds of topological information, i.e., low-pass and high-pass filter capture homophilic and heterophilic structures, respectively. Consequently, our mixed filter can handle two types of graphs simultaneously.

  Moreover, similarity variation patterns with the mixed filter of $S$ and $A$ are similar. This means that $S$ does not loss too much topological information from $A$. Importantly, the similarity with low-pass or high-pass filter converges to larger values, which indicates that filters with $S$ and $H$ get the similar nodes close.

  

	\section{Conclusion}
	In this paper, we make the first attempt to address the challenge of node clustering without any prior knowledge of graph homophily. We design a mixed filter to jointly explore low-frequency and high-frequency components  by reconstructing two graphs from the raw graph, which can be generalized to other tasks. Moreover, to reduce the potential coupling between attribute and topology structure of data, we project the smoothed attribute and raw structure into two subspaces via two individual MLPs. Comprehensive experimental results on 11 graph benchmarks demonstrate the proposed method's impressive performance. Therefore, our method generalizes well to real data. One potential limitation is that there could be information loss resulting from feature reconstruction without considering structure. In future work, we aim to design a new architecture to handle this issue. 

	
	\appendix

	
	
	\section*{Acknowledgments}
	This work was supported by the National Natural Science Foundation
of China (No. 62276053).
	
        \bibliography{refs}

\begin{thebibliography}{48}
\providecommand{\natexlab}[1]{#1}
\providecommand{\url}[1]{\texttt{#1}}
\expandafter\ifx\csname urlstyle\endcsname\relax
  \providecommand{\doi}[1]{doi: #1}\else
  \providecommand{\doi}{doi: \begingroup \urlstyle{rm}\Url}\fi

\bibitem[Abu-El-Haija et~al.(2019)Abu-El-Haija, Perozzi, Kapoor, Alipourfard,
  Lerman, Harutyunyan, Ver~Steeg, and Galstyan]{abu2019mixhop}
Abu-El-Haija, S., Perozzi, B., Kapoor, A., Alipourfard, N., Lerman, K.,
  Harutyunyan, H., Ver~Steeg, G., and Galstyan, A.
\newblock Mixhop: Higher-order graph convolutional architectures via sparsified
  neighborhood mixing.
\newblock In \emph{international conference on machine learning}, pp.\  21--29.
  PMLR, 2019.

\bibitem[Bi et~al.(2022)Bi, Du, Fu, Wang, Han, and Zhang]{DHGR}
Bi, W., Du, L., Fu, Q., Wang, Y., Han, S., and Zhang, D.
\newblock Make heterophily graphs better fit gnn: A graph rewiring approach.
\newblock \emph{arXiv preprint arXiv:2209.08264}, 2022.

\bibitem[Bo et~al.(2020)Bo, Wang, Shi, Zhu, Lu, and Cui]{SDCN}
Bo, D., Wang, X., Shi, C., Zhu, M., Lu, E., and Cui, P.
\newblock Structural deep clustering network.
\newblock In \emph{Proceedings of The Web Conference 2020}, pp.\  1400--1410,
  2020.

\bibitem[Bo et~al.(2021)Bo, Wang, Shi, and Shen]{FAGCN}
Bo, D., Wang, X., Shi, C., and Shen, H.
\newblock Beyond low-frequency information in graph convolutional networks.
\newblock In \emph{Thirty-Fifth {AAAI} Conference on Artificial Intelligence,
  {AAAI} 2021}, pp.\  3950--3957. {AAAI} Press, 2021.

\bibitem[Chen et~al.(2020)Chen, Wei, Huang, Ding, and Li]{GCNII}
Chen, M., Wei, Z., Huang, Z., Ding, B., and Li, Y.
\newblock Simple and deep graph convolutional networks.
\newblock In \emph{International Conference on Machine Learning}, pp.\
  1725--1735. PMLR, 2020.

\bibitem[Cheng et~al.(2021)Cheng, Wang, Tao, Xie, and Gao]{MAGCN}
Cheng, J., Wang, Q., Tao, Z., Xie, D., and Gao, Q.
\newblock Multi-view attribute graph convolution networks for clustering.
\newblock In \emph{Proceedings of the Twenty-Ninth International Conference on
  International Joint Conferences on Artificial Intelligence}, pp.\
  2973--2979, 2021.

\bibitem[Chien et~al.(2021)Chien, Peng, Li, and Milenkovic]{GPR}
Chien, E., Peng, J., Li, P., and Milenkovic, O.
\newblock Adaptive universal generalized pagerank graph neural network.
\newblock In \emph{9th International Conference on Learning Representations,
  {ICLR} 2021,}, 2021.

\bibitem[Cui et~al.(2020)Cui, Zhou, Yang, and Liu]{AGE}
Cui, G., Zhou, J., Yang, C., and Liu, Z.
\newblock Adaptive graph encoder for attributed graph embedding.
\newblock In \emph{Proceedings of the 26th ACM SIGKDD International Conference
  on Knowledge Discovery \& Data Mining}, pp.\  976--985, 2020.

\bibitem[Fan et~al.(2020)Fan, Wang, Shi, Lu, Lin, and Wang]{O2MAC}
Fan, S., Wang, X., Shi, C., Lu, E., Lin, K., and Wang, B.
\newblock One2multi graph autoencoder for multi-view graph clustering.
\newblock In \emph{Proceedings of The Web Conference 2020}, pp.\  3070--3076,
  2020.

\bibitem[Fang et~al.(2022)Fang, Wen, Kang, and Liu]{fang2022structure}
Fang, R., Wen, L., Kang, Z., and Liu, J.
\newblock Structure-preserving graph representation learning.
\newblock In \emph{Proceedings of the IEEE International Conference on Data
  Mining (ICDM)}, 2022.

\bibitem[Hassani \& Khasahmadi(2020)Hassani and Khasahmadi]{MVGRL}
Hassani, K. and Khasahmadi, A.~H.
\newblock Contrastive multi-view representation learning on graphs.
\newblock In \emph{International Conference on Machine Learning}, pp.\
  4116--4126. PMLR, 2020.

\bibitem[He et~al.(2021)He, Wei, Xu, et~al.]{BernNet}
He, M., Wei, Z., Xu, H., et~al.
\newblock Bernnet: Learning arbitrary graph spectral filters via bernstein
  approximation.
\newblock \emph{Advances in Neural Information Processing Systems},
  34:\penalty0 14239--14251, 2021.

\bibitem[Hou et~al.(2022)Hou, Liu, Dong, Wang, Tang, et~al.]{GraphMAE}
Hou, Z., Liu, X., Dong, Y., Wang, C., Tang, J., et~al.
\newblock Graphmae: Self-supervised masked graph autoencoders.
\newblock \emph{SIGKDD}, 2022.

\bibitem[Huang et~al.(2019)Huang, Frederking, et~al.]{RWR}
Huang, P.-Y., Frederking, R., et~al.
\newblock Rwr-gae: Random walk regularization for graph auto encoders.
\newblock \emph{arXiv preprint arXiv:1908.04003}, 2019.

\bibitem[Kang et~al.(2022)Kang, Liu, Pan, and Tian]{FGC}
Kang, Z., Liu, Z., Pan, S., and Tian, L.
\newblock Fine-grained attributed graph clustering.
\newblock In \emph{Proceedings of the 2022 SIAM International Conference on
  Data Mining (SDM)}, pp.\  370--378. SIAM, 2022.

\bibitem[Kipf \& Welling()Kipf and Welling]{GAE}
Kipf, T.~N. and Welling, M.
\newblock Variational graph auto-encoders.
\newblock \emph{Bayesian Deep Learning Workshop (NIPS 2016)}.

\bibitem[Li et~al.(2022)Li, Zhu, Cheng, Shan, Luo, Li, and Qian]{GloGNN}
Li, X., Zhu, R., Cheng, Y., Shan, C., Luo, S., Li, D., and Qian, W.
\newblock Finding global homophily in graph neural networks when meeting
  heterophily.
\newblock In \emph{International Conference on Machine Learning, {ICML} 2022},
  volume 162, pp.\  13242--13256. {PMLR}, 2022.

\bibitem[Lim et~al.(2021{\natexlab{a}})Lim, Hohne, Li, Huang, Gupta, Bhalerao,
  and Lim]{lim2}
Lim, D., Hohne, F., Li, X., Huang, S.~L., Gupta, V., Bhalerao, O., and Lim,
  S.~N.
\newblock Large scale learning on non-homophilous graphs: New benchmarks and
  strong simple methods.
\newblock \emph{Advances in Neural Information Processing Systems},
  34:\penalty0 20887--20902, 2021{\natexlab{a}}.

\bibitem[Lim et~al.(2021{\natexlab{b}})Lim, Li, Hohne, and Lim]{lim1}
Lim, D., Li, X., Hohne, F., and Lim, S.-N.
\newblock New benchmarks for learning on non-homophilous graphs.
\newblock \emph{International World Wide Web Conferences}, 2021{\natexlab{b}}.

\bibitem[Lin \& Kang(2021)Lin and Kang]{MvAGC}
Lin, Z. and Kang, Z.
\newblock Graph filter-based multi-view attributed graph clustering.
\newblock In \emph{IJCAI}, pp.\  2723--2729, 2021.

\bibitem[Lin et~al.(2023)Lin, Kang, Zhang, and Tian]{lin2023multi}
Lin, Z., Kang, Z., Zhang, L., and Tian, L.
\newblock Multi-view attributed graph clustering.
\newblock \emph{IEEE Transactions on Knowledge and Data Engineering},
  35\penalty0 (2):\penalty0 1872--1880, 2023.

\bibitem[Liu et~al.(2021)Liu, Wen, Kang, Luo, and Tian]{liu2021self}
Liu, C., Wen, L., Kang, Z., Luo, G., and Tian, L.
\newblock Self-supervised consensus representation learning for attributed
  graph.
\newblock In \emph{Proceedings of the 29th ACM International Conference on
  Multimedia}, pp.\  2654--2662, 2021.

\bibitem[Liu et~al.(2022{\natexlab{a}})Liu, Kang, Ruan, and
  He]{liu2022multilayer}
Liu, L., Kang, Z., Ruan, J., and He, X.
\newblock Multilayer graph contrastive clustering network.
\newblock \emph{Information Sciences}, 613:\penalty0 256--267,
  2022{\natexlab{a}}.

\bibitem[Liu et~al.(2022{\natexlab{b}})Liu, Tu, Zhou, Liu, Song, Yang, and
  Zhu]{DCRN}
Liu, Y., Tu, W., Zhou, S., Liu, X., Song, L., Yang, X., and Zhu, E.
\newblock Deep graph clustering via dual correlation reduction.
\newblock In \emph{Proc. of AAAI}, 2022{\natexlab{b}}.

\bibitem[Luan et~al.(2021)Luan, Hua, Lu, Zhu, Zhao, Zhang, Chang, and
  Precup]{ACMNN}
Luan, S., Hua, C., Lu, Q., Zhu, J., Zhao, M., Zhang, S., Chang, X.-W., and
  Precup, D.
\newblock Is heterophily a real nightmare for graph neural networks to do node
  classification?
\newblock \emph{arXiv preprint arXiv:2109.05641}, 2021.

\bibitem[Ma et~al.(2022)Ma, Liu, Shah, and Tang]{HM}
Ma, Y., Liu, X., Shah, N., and Tang, J.
\newblock Is homophily a necessity for graph neural networks?
\newblock In \emph{The Tenth International Conference on Learning
  Representations, {ICLR} 2022}, 2022.

\bibitem[Mrabah et~al.(2022)Mrabah, Bouguessa, Touati, and Ksantini]{UATEatBat}
Mrabah, N., Bouguessa, M., Touati, M.~F., and Ksantini, R.
\newblock Rethinking graph auto-encoder models for attributed graph clustering.
\newblock \emph{IEEE Transactions on Knowledge and Data Engineering}, 2022.

\bibitem[Ning et~al.(2022)Ning, Wang, Wang, Qiao, Fan, Zhang, Du, and
  Zhou]{hop}
Ning, Z., Wang, P., Wang, P., Qiao, Z., Fan, W., Zhang, D., Du, Y., and Zhou,
  Y.
\newblock Graph soft-contrastive learning via neighborhood ranking.
\newblock \emph{arXiv preprint arXiv:2209.13964}, 2022.

\bibitem[Pan \& Kang(2021)Pan and Kang]{MCGC}
Pan, E. and Kang, Z.
\newblock Multi-view contrastive graph clustering.
\newblock \emph{Advances in neural information processing systems},
  34:\penalty0 2148--2159, 2021.

\bibitem[Pan et~al.(2019)Pan, Hu, Fung, Long, Jiang, and Zhang]{ARVGA}
Pan, S., Hu, R., Fung, S.-f., Long, G., Jiang, J., and Zhang, C.
\newblock Learning graph embedding with adversarial training methods.
\newblock \emph{IEEE transactions on cybernetics}, 50\penalty0 (6):\penalty0
  2475--2487, 2019.

\bibitem[Pei et~al.(2020)Pei, Wei, Chang, Lei, and Yang]{Geom-GCN}
Pei, H., Wei, B., Chang, K.~C., Lei, Y., and Yang, B.
\newblock Geom-gcn: Geometric graph convolutional networks.
\newblock In \emph{8th International Conference on Learning Representations,
  {ICLR} 2020,}, 2020.

\bibitem[Peng et~al.(2021)Peng, Liu, Jia, and Hou]{AGN-H}
Peng, Z., Liu, H., Jia, Y., and Hou, J.
\newblock Attention-driven graph clustering network.
\newblock In \emph{Proceedings of the 29th ACM International Conference on
  Multimedia}, pp.\  935--943, 2021.

\bibitem[Rozemberczki et~al.(2021)Rozemberczki, Allen, and Sarkar]{squirel}
Rozemberczki, B., Allen, C., and Sarkar, R.
\newblock Multi-scale attributed node embedding.
\newblock \emph{Journal of Complex Networks}, 9\penalty0 (2), 2021.

\bibitem[Topping et~al.(2022)Topping, Di~Giovanni, Chamberlain, Dong, and
  Bronstein]{SDRF}
Topping, J., Di~Giovanni, F., Chamberlain, B.~P., Dong, X., and Bronstein,
  M.~M.
\newblock Understanding over-squashing and bottlenecks on graphs via curvature.
\newblock In \emph{International Conference on Learning Representations}, 2022.

\bibitem[Tu et~al.(2021)Tu, Zhou, Liu, Guo, Cai, Zhu, and Cheng]{DFCN}
Tu, W., Zhou, S., Liu, X., Guo, X., Cai, Z., Zhu, E., and Cheng, J.
\newblock Deep fusion clustering network.
\newblock In \emph{Proceedings of the AAAI Conference on Artificial
  Intelligence}, volume~35, pp.\  9978--9987, 2021.

\bibitem[Wang et~al.(2019)Wang, Pan, Hu, Long, Jiang, and Zhang]{DAEGC}
Wang, C., Pan, S., Hu, R., Long, G., Jiang, J., and Zhang, C.
\newblock Attributed graph clustering: {A} deep attentional embedding approach.
\newblock In \emph{Proceedings of the Twenty-Eighth International Joint
  Conference on Artificial Intelligence}, pp.\  3670--3676, 2019.

\bibitem[Wang et~al.(2021)Wang, Wu, Zhang, Zhou, Chen, and Liu]{MSGA}
Wang, T., Wu, J., Zhang, Z., Zhou, W., Chen, G., and Liu, S.
\newblock Multi-scale graph attention subspace clustering network.
\newblock \emph{Neurocomputing}, 459:\penalty0 302--314, 2021.

\bibitem[Wang et~al.(2022)Wang, Jin, Wang, He, and Huang]{HOG-GNN}
Wang, T., Jin, D., Wang, R., He, D., and Huang, Y.
\newblock Powerful graph convolutional networks with adaptive propagation
  mechanism for homophily and heterophily.
\newblock In \emph{Proceedings of the AAAI Conference on Artificial
  Intelligence}, volume~36, pp.\  4210--4218, 2022.

\bibitem[Wang \& Zhang(2022)Wang and Zhang]{JacobiNet}
Wang, X. and Zhang, M.
\newblock How powerful are spectral graph neural networks.
\newblock In \emph{International Conference on Machine Learning, {ICML} 2022,},
  volume 162, pp.\  23341--23362. {PMLR}, 2022.

\bibitem[Wang \& Derr(2021)Wang and Derr]{TDGCN}
Wang, Y. and Derr, T.
\newblock Tree decomposed graph neural network.
\newblock In \emph{Proceedings of the 30th ACM International Conference on
  Information \& Knowledge Management}, pp.\  2040--2049, 2021.

\bibitem[Xia et~al.(2021)Xia, Gao, Yang, and Gao]{SAGC}
Xia, W., Gao, Q., Yang, M., and Gao, X.
\newblock Self-supervised contrastive attributed graph clustering.
\newblock \emph{arXiv preprint arXiv:2110.08264}, 2021.

\bibitem[Xie et~al.(2023)Xie, Chen, Kang, and Peng]{xie2023contrastive}
Xie, X., Chen, W., Kang, Z., and Peng, C.
\newblock Contrastive graph clustering with adaptive filter.
\newblock \emph{Expert Systems with Applications}, 219:\penalty0 119645, 2023.

\bibitem[Yang et~al.(2021)Yang, Li, Liu, Niu, Wang, Cao, and Guo]{DMP}
Yang, L., Li, M., Liu, L., Niu, B., Wang, C., Cao, X., and Guo, Y.
\newblock Diverse message passing for attribute with heterophily.
\newblock In \emph{Advances in Neural Information Processing Systems 34,
  NeurIPS 2021,}, pp.\  4751--4763, 2021.

\bibitem[Zbontar et~al.(2021)Zbontar, Jing, Misra, LeCun, and
  Deny]{BarlowTwins}
Zbontar, J., Jing, L., Misra, I., LeCun, Y., and Deny, S.
\newblock Barlow twins: Self-supervised learning via redundancy reduction.
\newblock In \emph{International Conference on Machine Learning}, pp.\
  12310--12320. PMLR, 2021.

\bibitem[Zhang et~al.(2019)Zhang, Liu, Li, and Wu]{AGC}
Zhang, X., Liu, H., Li, Q., and Wu, X.-M.
\newblock Attributed graph clustering via adaptive graph convolution.
\newblock pp.\  4327--4333, 2019.

\bibitem[Zhu \& Koniusz(2021)Zhu and Koniusz]{S2GC}
Zhu, H. and Koniusz, P.
\newblock Simple spectral graph convolution.
\newblock In \emph{9th International Conference on Learning Representations,
  {ICLR} 2021,}, 2021.

\bibitem[Zhu et~al.(2020)Zhu, Yan, Zhao, Heimann, Akoglu, and Koutra]{H2GCN}
Zhu, J., Yan, Y., Zhao, L., Heimann, M., Akoglu, L., and Koutra, D.
\newblock Beyond homophily in graph neural networks: Current limitations and
  effective designs.
\newblock \emph{Advances in Neural Information Processing Systems},
  33:\penalty0 7793--7804, 2020.

\bibitem[Zhu et~al.(2022)Zhu, Li, Wang, Xiao, Zhao, and Hu]{CDRS}
Zhu, P., Li, J., Wang, Y., Xiao, B., Zhao, S., and Hu, Q.
\newblock Collaborative decision-reinforced self-supervision for attributed
  graph clustering.
\newblock \emph{IEEE Transactions on Neural Networks and Learning Systems},
  2022.

\end{thebibliography}
        \bibliographystyle{icml2023}
	
\end{document}